\documentstyle[aps,12pt,epsf,psfig]{revtex}
\begin{document}

\draft

\title{
Thermal dilepton signal versus dileptons from open charm and
bottom decays in heavy-ion collisions
}

\author{
{\sc
K. Gallmeister$^a$, B. K\"ampfer$^a$, O.P. Pavlenko$^{a,b}$}
}

\address{
$^a$Research Center Rossendorf, PF 510119, 01314 Dresden, Germany \\[1mm]
$^b$Institute for Theoretical Physics, 252143 Kiev - 143, Ukraine
}

\maketitle

\vspace*{1cm}

\begin{abstract}
We analyze the opportunity to observe thermal dileptons emitted off
deconfined matter resulting in heavy-ion collisions at RHIC and LHC energies.
Special kinematical conditions provided by the detector systems
PHENIX and ALICE,
and the so-called $M_\perp$ scaling behavior of thermal dilepton spectra
are taken into account.
Our considerations include energy loss effects of the fast heavy quarks in
deconfined matter, which for themselves can help to identify the creation
of a hot and dense parton medium. 
Due to a threshold like effect for decay leptons we find a window
at large transverse pair momentum and fixed transverse mass
where the thermal signal can exceed the background of
dileptons from correlated semileptonic decays of charm and bottom mesons.
\end{abstract}


\newpage

\section{Introduction}

One of the ultimate goals in investigating central heavy-ion collisions
at very
high energies is to analyze the properties of highly excited and
deconfined matter. Direct probes, like electromagnetic signals
\cite{Shuryak1}, are considered
as useful carriers of nearly undisturbed information
from the transient hot reaction stages. Pairs of electrons and muons
are experimentally accessible, and will be measured also in forthcoming
heavy-ion experiments \cite{PHENIX_prop,ALICE_prop}.
These types of probes have proven useful
at Super Proton Synchrotron (SPS) energies at CERN
in the resonance region below the $\omega$, $\rho$
\cite{CERES}, and in the $J/\psi$ region
\cite{J_psi}; also in the so-called continuum region in between
a yet unexplained excess seems to be found \cite{continuum}.
However, recent estimates \cite{Vogt1}
of dileptons stemming from semileptonic decays of open charm
and bottom mesons produced in the same central collisions of heavy ions
show that these represent the dominating
dilepton source at energies envisaged at the
Relativistic Heavy Ion Collider (RHIC) in Brookhaven and
at the Large Hadron Collider (LHC) in CERN.
In particular, the dilepton signal from thermalized,
strongly interacting matter
in the intermediate reaction stages is estimated to be up to
two orders of magnitude below the background of correlated decays
of open charm or bottom meson pairs in a wide invariant mass region.
Even strong energy loss effects of the heavy quarks in the hot
and dense medium \cite{Shuryak,PLB98,Lin} do not basically change this
situation: The decay dileptons might be suppressed down to the Drell-Yan
background, but it is questionable whether such strong energy loss effects
can happen in reality \cite{Baier_new}. In this situation at least two
issues can be seen.
First, it is necessary to extend the traditional strategy
for searching a thermal dilepton signal from deconfined matter.
The usual approach relies on the analysis of the invariant mass ($M$) spectra.
One can expect, however, that the double differential dilepton spectrum
as a function of the transverse pair momentum $Q_\perp$
and transverse pair mass $M_\perp = \sqrt{M^2 + Q_\perp^2}$
contains much more information
which can be used to impose special kinematical gates for finding
a window to observe the thermal dileptons in the continuum region
in spite of the large decay background.
Second, one can try to employ the initial hard production of open charm
and bottom followed by semileptonic decays to probe deconfined matter
by energy loss effects. A similar idea for hard jet production is
widely under discussion now \cite{Wang}.

In the present paper we focus on both of the above issues.
We compare the so-called $M_\perp$ scaling property of the dielectrons
stemming from deconfined matter with
pairs resulting from correlated open charm and bottom decays.
As well-known, under certain conditions the dilepton yield from a thermalized
quark-gluon plasma depends only on the transverse mass, i.e.,
there is no dependence on the variable $Q_\perp$ at fixed $M_\perp$
in a given rapidity interval $Y$.
At the same time the corresponding $Q_\perp$ dependence of dileptons from
heavy quark decays has qualitatively a different behavior: It is peaked at
small values of $Q_\perp$ and drops rapidly at increasing values of
the transverse momentum.
To get realistic estimates of the $M_\perp$ scaling we perform Monte Carlo
simulations for both the correlated heavy quark decays and the
thermal production including the acceptance of the detectors
PHENIX at RHIC and ALICE at LHC.
To strengthen the $M_\perp$ scaling violation for dileptons from
decays we select only such leptons with individual transverse
momentum $p_\perp >  p_\perp^{\rm min}$ with appropriate value
of $p_\perp^{\rm min}$.
We find that, while the thermal dilepton yield from the plasma is less
affected by the kinematical restriction,
the contributions from decays at large transverse pair momentum
is reduced drastically. Actually
this leads to a threshold like behavior of the transverse momentum spectrum
at fixed value of $M_\perp$ for dileptons from heavy quark decays
in contrast to the approximate plateau of the thermal yield which
extends up to the kinematical limit at $Q_\perp = M_\perp$.
As a result, an opportunity is found which allows to separate the thermal
dilepton signal from the background of heavy quark decays on the basis
of the $M_\perp$ scaling property.
We also study to what extent the single electron and dilepton spectra
are modified
by the energy loss effects of heavy quarks when traversing deconfined matter.
In doing so we concentrate on calculable processes in the first
few fm/c of the collision course. There are, of course, other processes
in the later reaction stages that might considerably affect the momenta of the
charm and bottom mesons \cite{Svetitsky}, and also thermal radiation
of semihard probes can stem from later stages. For instance,
in ref.~\cite{Santorin} the hypothesis has been formulated, resting
on an analysis of present Pb + Pb data at SPS energy, that all hadron
spectra can be parametrized by a common flow velocity
and freeze-out temperature.
This, if correct for higher energies too, would imply that charm and
bottom spectra are much stronger modified than assumed in the following,
and as a consequence the chances to identify a thermal dilepton source would
be better.

Our paper is organized as follows. In section 2 we present our calculation
procedure of charm and bottom spectra and the resulting single electron
and dielectron spectra. Section 3 is devoted to the thermal dileptons
at midrapidity
and the $M_\perp$ scaling properties. Our conclusions can be found
in section 4.

\section{Dileptons from open charm and bottom decays}

\subsection{Initial charm and bottom production}

In our approach we utilize the leading order QCD processes
$gg \to Q \bar Q$ and $q \bar q \to Q \bar Q$ for heavy quark - antiquark
($Q \bar Q$)
production and simulate higher order corrections by an appropriate
${\cal K}_Q$ factor. We have checked that
such a procedure reproduces within the needed
accuracy the more involved next-to-leading order calculations \cite{Vogt1}.
The number of $Q \bar Q$ pairs, produced initially with transverse
momenta $p_{\perp 1} = - p_{\perp 2} = p_\perp$ at rapidities $y_{1,2}$
in central $AA$ collisions can be calculated by
\begin{eqnarray} \label{eq.1}
dN_{Q \bar Q}
& = &
T_{AA}(0) \, {\cal K}_Q \, H(y_1,y_2,p_\perp) \,
dp_\perp^2 \, dy_1 \, dy_2, \\
H(y_1,y_2,p_\perp)
& = &
x_1 \, x_2 \left\{ f_g(x_1,\hat Q^2) \, f_g(x_2,\hat Q^2)
\frac{d \hat \sigma_g^Q}{d \hat t} \right. \nonumber \\
&& +
\left. \sum_{q \bar q} \left[
f_q(x_1,\hat Q^2) \, f_{\bar q} (x_2,\hat Q^2) +
f_q(x_2,\hat Q^2) \, f_{\bar q} (x_1,\hat Q^2)\right]
\frac{d \hat \sigma_q^Q}{d \hat t} \right\},
\end{eqnarray}
where $\hat \sigma^Q_{q,g}/d \hat t$ are
elementary cross sections (see for details \cite{Vogt1,PLB97}),
$f_i(x,\hat Q^2)$ with $i=g,q,\bar q$
denote the parton structure functions,
$x_{1,2} = m_\perp \left(
\exp\{ \pm y_1 \} + \exp\{ \pm y_2 \}  \right) /\sqrt{s}$ and
$m_\perp = \sqrt{p_\perp^2 + m_Q^2}$.
As heavy quark masses we take $m_c =$ 1.5 GeV and $m_b =$ 4.5 GeV.
We employ the HERA supported structure function set MRS D'- \cite{MRS}
from the PDFLIB at CERN.
Nuclear shadowing effects are not included since they will be considered
separately. In addition, for our present goals we expect no significant
modification by shadowing in the large transverse momentum region
according to results of ref.~\cite{Vogt2}.
The overlap function for central collisions is
$T_{AA}(0) = A^2/(\pi R_A^2)$
with  $R_A = 1.1 A^{1/3}$ fm and $A = 200$.
From a comparison with results of ref.~\cite{Vogt1}
we find the scale $\hat Q^2 = 4 M_Q^2$ and ${\cal K}_Q =$ 2 as most
appropriate.

\subsection{Energy loss effects}

Energy loss effects
(cf. \cite{Baier_new,Srivastava,Baier,Baier2,Zakharov}
and further references therein) have been found important
\cite{Shuryak,PLB98,Lin}
since the momenta of back-to-back moving heavy quarks are degraded and the
resulting dileptons in correlated semileptonic decays get less invariant
mass. This process causes a considerable reduction of the number of
high invariant
mass dileptons from correlated $D \bar D$ and $B \bar B$ decays.

To model the energy loss effects of heavy quarks in expanding matter we assume
that a heavy quark has the same rapidity as the longitudinal flow
at the space-time point of the initial quark production.
Therefore, with respect to the fluid's local rest frame
the heavy quark has essentially only a transverse momentum $p_\perp$
which may depend on the proper local
time $\tau$ in accordance with the energy loss $dE/dx$
in transverse direction.
This picture implies a very fast thermalization of gluons and light quarks.

Unfortunately a sufficiently general treatment of the energy loss rate
$dE/dx$ for heavy quarks passing through expanding QCD matter has not
been developed so far.
Recently the total energy loss of a high-energy parton propagating
transversely through expanding deconfined matter has been derived
as $dE / dx |_{\rm expanding} = \xi dE / dx |_{T_f}$
\cite{Baier_new},
supposed the medium cools according to a power law and the
initial time is short. The numerical factor $\xi$ is
${\cal O}(2)$ for a parton created inside the medium.
Within such a scenario the stopping power depends
only on the final temperature at which the deconfined medium is left,
as indicated by the subscript $T_f$.
We denote this stopping scenario as model I and
take $dE / dx |_{T_f}$ from \cite{Baier_new,Baier,Baier2,Zakharov}.
This model is appropriate for LHC conditions, where the initial temperature
is very high and where the life time of deconfined matter is large enough
that most energetic heavy quarks can leave this stage before $T_c$
is reached.
For RHIC conditions we adopt the results of ref.~\cite{Baier}
for $dE/dx$ in a static medium but apply them in the simplest generalization
to a medium with time dependent density. This gives evolution equations
for the transverse momentum $p_\perp(\tau)$ of quarks propagating
the distance $r_\perp(\tau)$ in the transverse direction. The scenario
is denoted
as model II, which is described in more detail in refs.~\cite{PLB98,Santorin}.
Strictly speaking, due to the finite size of the medium, the large mass
of the heavy quarks and other approximations inherent in such approaches,
the models I and II provide at best semi-quantitative estimates.
Nevertheless one can hope to cover various limiting
cases to get some insight in the importance of the energy loss effects.

To get the spectra of charm and bottom quarks
after energy loss we use a Monte Carlo
simulation with a uniform distribution of the random initial position
and random orientation of $Q \bar Q$
pairs in the transverse plane.
In model I the total energy loss is determined by the transverse distance
$d$ of a created heavy quark to the boundary of the system, the quark
initial energy and the temperature when leaving the system.
Most quarks experience an energy loss $\propto d$ and only a few ones
$\propto d\,^2$.
(The transverse expansion at early times can be neglected,
and we take as averaged transverse radius of parton matter $R = 7$ fm.)
In model II we integrate the evolution equations for
the heavy quark momentum $p_\perp( \tau)$ and
position $r_\perp(\tau)$
together with the time evolution of the temperature
$T(\tau)$ and fugacities $\lambda_{q,g}(\tau)$ (see below)
as described in refs.~\cite{PLB98,PLB97}.
The quantity $r_\perp(\tau)$ is used to check
whether the considered heavy quark propagates still within
deconfined matter; if it leaves the deconfined medium it does not longer
experience an energy loss.
(The energy loss in a mixed and hadron phase might
be incorporated in line with the approach of \cite{Svetitsky}.)
We consider a heavy quark to be thermalized if its transverse mass
$m_\perp$
during energy loss becomes less than the averaged transverse mass
of thermalized light partons at given temperature.
These heavy quarks are redistributed according to the
transverse mass distribution
$dN_Q / dm_\perp \propto m_\perp^2 \exp\{ -m_\perp / T_{\rm eff} \}$
with an effective temperature $T_{\rm eff} = 150$ MeV \cite{Lin}.

\subsection{Single electron spectra}

We employ a delta function like fragmentation scheme for the heavy quark
conversion into $D$ and $B$ mesons. It results in the same transverse
momentum of the heavy meson as the heavy parent quark has had previously.
Inclusive single transverse momentum spectra of electrons/positrons
from semileptonic open charm and bottom decays, i.e.,
$D (B) \to e^+ X$,
and dilepton spectra from correlated decays as well,
i.e., $D (B) \bar D (\bar B) \to e^+ X e^- \bar X$,
are obtained from a Monte Carlo code which utilizes the inclusive
primary electron energy distribution as delivered by
JETSET~7.4. The heavy mesons are randomly decayed in their rest system
and the resulting electrons then boosted appropriately.
The average branching ratio of $D (B) \to e^+ X$
is taken as 12 (10)\% \cite{PDB}. The neglect of secondary electrons
is justified since we consider here single electron spectra
with $p_\perp >$ 1 GeV and dilepton spectra with $M >$ 2 GeV.
The average energies of the secondary electrons, in particular from
$B$ decays, are too small to
affect the spectra in these kinematical regions noticeable.

The following acceptance cuts are utilized in our Monte Carlo calculations.
PHENIX \cite{PHENIX_prop} can register electrons in the pseudorapidity window
$| \eta | \le 0.35$ with azimuthal coverage 0 - 90 and 135 - 225
degrees, while ALICE \cite{ALICE_prop} is azimuthally symmetric and covers
$| \eta | \le 0.9$. As minimum transverse electron momentum
we use $p_\perp =$ 1 GeV. A maximum transverse momentum cut is not
imposed, even at ALICE there is probably one.
We consider here only the electron and $e^+ e^-$ channels and the
midrapidity region since the energy loss effects are strongest there
\cite{Lin}.

The results of our calculations of the transverse momentum spectra
of inclusive single electrons at
RHIC conditions are depicted in figs.~1a and b. As already pointed out
in ref.~\cite{Tannenbaum}, a measurement of single electrons
with $p_\perp >$ 1.5 GeV at RHIC
can give a clean and background-free charm signal in heavy-ion collisions. Energy loss
effects of the heavy quarks modify this signal. As seen in figs.~1a and b
such modifications, according to the energy loss described above,
are measurable. For both the charm and the bottom decay the
transverse single electron spectra become softer.
This leads to a noticeable change of the slope of the spectra.
One can fit the distributions by
$d N_{e^-}/d p_\perp \propto \exp \{- p_\perp / T_e \} $
in the interval 2.5 GeV $< p_\perp <$ 4.5 GeV and finds
$T_e =$ 646 (512) MeV in case of no stopping (stopping according to
model II) for charm, and $T_e =$ 851 (640) MeV for bottom.
The superposition of charm and bottom decay spectra can be parametrized by
$T_e =$ 711 (574) MeV.
Therefore, in comparing the $p_\perp$
spectra of single electrons in pp and AA collisions at the same
beam energies one can reveal the stopping effect. (A direct measurement
of the $D (B)$ $p_\perp$ distributions would be more useful,
but it is unclear whether the $D (B)$ mesons
can be identified via their hadronic
decay modes in such high-multiplicity environments.)
In extreme case, when the stopping is so large that thermalization with the
surrounding matter happens, then a pronounced ''thermal bump'' at
$p_\perp <$ 2.5 GeV appears. All of these considerable modifications
of the transverse electron spectra in AA collisions in comparison with pp
collisions can be helpful for identifying the creation of a hot and dense
parton system. It should be noted that the energy loss effects affect
the single-electron $p_\perp$ spectra much stronger than another
fragmentation scheme (e.g. the below described Peterson scheme) would do.

\subsection{Decay dielectrons}

The invariant mass dilepton spectra resulting from
both correlated charm and bottom decays are displayed in figs.~2a and b
for RHIC and LHC energies. Also shown is the distribution of thermal
dileptons according to eq.~(9) below.
As pointed out earlier \cite{Vogt1} the $M$ spectra are by far
dominated by the dileptons from correlated open charm and bottom decays.
A strong energy loss according to the above models reduces the decay
contributions, so that
(i) at RHIC charm and bottom decay dileptons are in the same order of
magnitude as the Drell-Yan background \cite{PLB98,Lin}, and
(ii) at LHC still bottom will dominate the Drell-Yan background, and
charm is slightly below bottom \cite{PLB98}.
While without any detector acceptances usually the predictions
of the thermal yield (dealt with in the next section)
is in the same order of magnitude as the Drell-Yan
background \cite{Vogt1,PLB98,PLB97,PRC95}, it turns out that
both the acceptances of PHENIX and ALICE suppress the thermal signal
stronger than the Drell-Yan background.
At the first sight such a situation looks quite unfavorable
for a measurement of the thermal signal even at such high energies
as achieved at RHIC and LHC. Nevertheless we show below that one can
select special kinematical conditions superimposed to the detector
cuts to analyze the more informative double differential dilepton
spectrum with the aim to find a window for the thermal signal.

\section{Thermal dilepton production and $M_\perp$ scaling}

\subsection{Spectra of thermal dileptons}

In order to estimate the thermal dilepton yield from deconfined matter
we restrict ourselves to the lowest order processes
$q \bar q \to \gamma^* \to e^+ e^-$, i.e., the electromagnetic
annihilation of quarks and antiquarks. Contributions from QCD processes
like $q \bar q \to g e^+ e^-$ and $ q g \to q e^+ e^-$ have been
considered \cite{PRC95} and appear to be not very significant
(typically they increase the yield by a factor less than 2.5).
The rate of dilepton production per space-time volume can be obtained
within a kinetic theory approach as
\begin{eqnarray}
\frac{dN_{e^+ e^-}}{d^4 x}
& = & \int
\frac{d^3p_1}{2E_1 (2\pi)^3}
\frac{d^3p_2}{2E_2 (2\pi)^3}
\frac{d^3p_+}{2E_+ (2\pi)^3}
\frac{d^3p_-}{2E_- (2\pi)^3} \,
f_1(x,p_1) \, f_2(x,p_2) \nonumber \\[2mm]
&& \times
|{\cal M}(12 \to e^+ e^-)|^2 \,
(2\pi)^4 \, \delta^{(4)} (p_1 + p_2 - p_+ - p_-),
\end{eqnarray}
where $p_a = (E_a, \vec p_a)$ are the four momenta of incoming quarks
($a =$ 1, 2) and outgoing leptons ($a = e^+, e^-$). The quantity
$|{\cal M}(12 \to e^+ e^-)|^2$ denotes the square of the amplitude
of the process, summed over initial and final state particles.
The quantities $f_a(x,p_a)$ stand for the distribution functions of quarks
and antiquarks. Integrating eq.~(3) over the momenta of the leptons one gets
a widely used expression \cite{Ruuskanen}
for the dilepton rate per four-momentum of the pair,
$Q = p_+ + p_-$,
\begin{equation}
\frac{dN_{e^+ e^-}}{d^4x \, d^4Q} = \int
\frac{d^3p_1}{(2\pi)^3}
\frac{d^3p_2}{(2\pi)^3} \,
f_1(x,p_1) \, f_2(x,p_2) \,
v \, \sigma (M^2) \,
\delta^{(4)} (p_1 + p_2 - Q).
\end{equation}
The pair's four momentum can be expressed as
$Q = (M_\perp \mbox{ch} Y, \vec Q_\perp, M_\perp \mbox{sh} Y)$
with $M_\perp = \sqrt{M^2 + Q_\perp^2}$ as transverse mass,
$M$ as invariant mass, and $Y$ as rapidity of the pair.
The relative quark velocity reads $v = M/(2 E_1 E_2)$, while the total
cross section is
$\sigma(M^2) = \frac{4 \pi \alpha^2}{3 M^2} 12 F_q$ with
quark charge factor
$F_q = \sum e_q^2 = \frac 59 $ for u,\,d quarks, and $\alpha$ is the
fine structure constant.

For the calculation of dilepton spectra in the region of large invariant
mass one can utilize the Boltzmann approximation for the incoming
partons
\begin{equation}
f_a(x,p) = \lambda_a(x) \exp
\left\{
\frac{p_a \cdot u(x)}{T(x)}
\right\},
\end{equation}
where $u$ is the four-velocity of the medium. The above parton
distributions are assumed to be in thermal equilibrium
in momentum space, but not necessarily in chemical equilibrium.
Only for chemical equilibrium the fugacities $\lambda_a$ become unity.

\subsection{$M_\perp$ scaling}

The space-time evolution of the produced deconfined matter is governed
in our approach by the longitudinal scaling-invariant  expansion
accompanied by quark and gluon chemical equilibration processes
\cite{PRC95,Biro}. Within such a model the space-time volume reads
$d^4 x = \pi R_A^2 \tau d \tau d \tilde \eta$ with $\tau$ as proper time
and flow rapidity $\tilde \eta$.
Integrating the rate eq.~(4) over the parton momenta and
space-time evolution results in the dilepton spectrum
\begin{equation}
\frac{dN_{e^+ e^-}}{dM_\perp^2 dQ_\perp^2 dY} =
\frac{\alpha^2 R_A^2}{4 \pi^2}
F_q
\int d \tau \tau
K_0 \left( \frac{M_\perp}{T(\tau)} \right)
\lambda_q^2(\tau).
\label{m_perp}
\end{equation}
The physical information encoded in the spectrum eq.~(\ref{m_perp})
can be inferred from the following approximation.
For large values of $M_\perp$ the main contribution to the time integral
stems from early, hot stages at $\tau \sim \tau_i$. Under the assumption
that the equation of state behaves like $p \propto e$ (here $p$ and $e$
stand for the pressure and energy density, respectively)
one can derive from energy-momentum conservation the relation
$T = T_i (\tau_i/\tau)^{1/3} \hat F(\tau)$,
$\hat F =[(d_g \lambda^g_i + d_q \lambda^q_i)/
(d_g \lambda^g(\tau) + d_q \lambda^q(\tau))]^{1/4}$
with $d_{q,g}$ as degeneracy factors of quarks and gluons.
Since near $\tau_i$ one can approximate $\hat F \approx 1$ and
$\lambda^q \approx \lambda^q_i$, one arrives at
\begin{equation}
\frac{d N_{e^+ e^-}}{dY \, d M_\perp^2 \, d Q_\perp^2}
\approx
\frac{3 \alpha^2 R_A^2  F_q}{4 \pi^2}
\left( \frac{\tau_i \lambda^q_i T_i^3}{M_\perp^3} \right)^2 \,
H \left( \frac{M_\perp}{T_i}\right)
\label{m_perp_a}
\end{equation}
with $H(x) = x^3 (8+x^2) \, K_3(x)$ and
$K_n$ as modified Bessel function of $n$th order.
For the possible initial conditions at LHC (see below) eq.~(\ref{m_perp_a})
represents a fairly good approximation, also for small values
of $\lambda^{q,g}_i$.
Eq.~(\ref{m_perp_a}) has the structure
$\hat f_1(\tau_i \lambda^q_i) \, \hat f_2(M_\perp/T_i)$, therefore one can
infer from it the value of $T_i$ by measuring the transverse rate
at two distinct values of $M_\perp$. Afterwards, the combination
$\tau_i \lambda^q_i$ can be extracted. If one could constrain by other
means the initial time of the thermalized era, $\tau_i$,
then even eq.~(\ref{m_perp_a}) allows to estimate the initial fugacity
$\lambda^q_i$. Due to particle production processes the comoving
entropy density does not longer serve as a link between
initial and final states in the evolution dynamics.

Our choice of initial conditions for produced deconfined matter is based
on the estimates of refs.~\cite{PLB97,Eskola1} for the minijet plasma with a
suitable parametrization of the soft component \cite{PLB98} which are
similar to the self-screened parton cascade model \cite{Eskola2}.
We take as main set of parameters for the initial temperature
$T_i =$ 550 (1000) MeV, for gluon fugacity
$\lambda_i^g =$ 0.5, and for light quark fugacity
$\lambda_i^q = \frac 15 \lambda_i^g$
of the parton plasma formed at RHIC (LHC) at initial time
$\tau_i =$ 0.2 fm/c.
To achieve an upper ''optimistic'' limit of the thermal dilepton yield
we consider below also higher values of the initial quark and gluon fugacities
($\lambda_i^q =$ 0.5, 1, and $\lambda_i^g =$ 1)
which lead to different final states.
For the sake of definiteness we assume
full saturation at confinement temperature $T_c =$ 170 MeV
and a quadratic time dependence of $\lambda^{q,g}(\tau)$ \cite{PLB97,PRC95}.

To get realistic dilepton spectra which can be measured in relativistic
heavy-ion collisions one should take into account the experimental conditions
related to the acceptance of detectors.
To do so we employ Monte Carlo simulations
with proper distributions of created leptons with respect to individual
rapidities $y_{\pm}$ and transverse momenta $\vec p_{\perp \pm}$.
In case of the thermal
emission of leptons from deconfined matter the needed distribution
can be obtained directly from eq.~(3). Due to the energy-momentum
conservation the exponential function, originating from the Boltzmann
distribution, can be rewritten as
$\exp \{ (p_1 + p_2)\cdot u /T \} = \exp \{ (p_+ + p_-)\cdot u /T\}$.
Since also
$|{\cal M} (12 \to e^+ e^-) |^2 = |{\cal M } (e^+ e^- \to 12) |^2$
the integration in eq.~(3) over the parton momenta yields
\begin{equation}
\frac{dN_{e^+ e^-}}{d^4x} = \int
\frac{d^3p_+}{2E_+ (2\pi)^3}
\frac{d^3p_-}{2E_- (2\pi)^3}
\exp \left\{ - \frac{(p_+ + p_-) \cdot u}{T} \right\}
F
\sigma(e^+ e^- \to 12),
\end{equation}
where $\sigma(e^+ e^- \to 12) = \sigma(12 \to e^+ e^-) = \sigma(M^2)$
and the flux factor becomes $F = 2 M^2$.
Then the space-time integrated distribution of thermal leptons reads
\begin{equation}
\frac{dN_{e^+ e^-}}{d^2p_+ \, d^2 p_- \, dy_+ \, dy_-} =
\frac{\alpha^2 R_A^2}{4 \pi^5} F_q
\int d \tau \tau K_0 \left( \frac{M_\perp}{T(\tau)} \right)
\lambda_q^2 (\tau),
\end{equation}
with dilepton transverse mass
$M_\perp^2 = p_{\perp+}^2 + p_{\perp-}^2 +
2 p_{\perp+} p_{\perp-} \mbox{ch} (y_+ - y_-)$.
To get the invariant mass spectra we also use the kinematical relationship
$M^2 =2 p_{\perp+} p_{\perp-} [\mbox{ch} (y_+ -y_-) - \cos(\phi_+ - \phi_-)]$,
where $\phi_{\pm}$ denote the azimuthal angles of the leptons
in the transverse plane.

As pointed out above,
due to the dominant contribution of correlated open charm and bottom decays
via $D (B) \bar D  (\bar B) \to e^+ X e^- \bar X$
into the high invariant mass dilepton spectra
it is very unlikely that the thermal dilepton production from
quark fusion can be observed by using the distribution
$dN_{e^+e^-} / dM^2$ alone even when imposing tricky cuts.
In this situation the more detailed information of the full spectrum
$dN_{e^+e^-} / dM_\perp^2 \, dQ_\perp^2 \, dY$ is expected to allow for
better chances to extract the wanted thermal signal.

As known since some time
\cite{PRC95,LMcL,Asakawa} and as directly seen in eq.~(6) the dilepton
yield from an equilibrium quark-gluon plasma depends only on the transverse
mass, i.e.,
$dN_{e^+e^-} / dM_\perp^2 \, dQ_\perp^2 \, dY \propto f(M_\perp)$;
that is, it scales with $M_\perp$.
The main conditions for this famous $M_\perp$ scaling are \cite{LMcL}
(i) local thermalization of the source of the dileptons,
(ii) no scale other then the temperature, and
(iii) predominant boost-invariant flow of matter.
The most convincing way to see the validity of the $M_\perp$
scaling is to consider the $Q_\perp$ dependence of the dilepton spectrum
at fixed value of $M_\perp$. In case of the boost-invariantly expanding
quark-gluon plasma one gets for
$dN_{e^+e^-} / dM_\perp^2 |_{M_\perp = \rm fix} \, dQ_\perp^2 \, dY$
indeed a straight line between
$Q_\perp =$ 0 and $M_\perp$. A detailed analysis of the transverse
momentum dependence of the dilepton spectra is performed in
ref.~\cite{PRC95}, where the possible origins of a $M_\perp$ scaling violation
related to transverse flow and contributions from the hadron gas are studied.
The transverse expansion of deconfined matter affects hardly the $M_\perp$
scaling, while for hadron matter the life time is strongly reduced, so that
its contribution to the total spectrum is moderate. In fact, our previous
studies \cite{PRC95} show that the conditions for the $M_\perp$ scaling
are satisfied for radiation from thermalized matter expected to be
created in central collisions at RHIC and LHC.
At the same time the dileptons from correlated open charm and bottom decay
should strongly violate the $M_\perp$ scaling if the $c \bar c$ and
$b \bar b$ pairs are dominantly produced by hard initial parton collisions.
The qualitative argument is as follows: Similar to the Drell-Yan process
the hard charm and bottom pairs are created mainly back-to-back
(unless hard gluon radiation spoils this correlation),
i.e., the pair momentum is preferential $Q_\perp^{Q \bar Q} \approx 0$,
and consequently also the $Q_\perp^{Q \bar Q}$ spectrum at fixed
$M_\perp^{Q \bar Q}$ is strongly
peaked at 0. The subsequent semileptonic decays cause a partial randomization
of the lepton directions but, due to the Lorentz boost effect,
both the leptons still are concentrated at small values of $Q_\perp$.
As a result one can expect that in the large $Q_\perp$ region the
thermal signal sticks out, while the semileptonic decay products
are sharply dropping.

\subsection{Monte Carlo simulations}

To quantify this effect we perform Monte Carlo calculations
generating thermal dielectrons in accordance with the distribution eq.~(9).
As above we take into account the acceptances of PHENIX at RHIC and ALICE
at LHC. First we consider the simplified case without any momentum
degradation of heavy quarks, i.e., energy loss effects are not included
and the $\delta$ function fragmentation scheme is employed.
This is obviously the most favorable case for the relative contribution from
decay dileptons to the full dilepton yield.

In order to get some threshold effect for the decay dileptons
in dependence of $Q_\perp$  at fixed $M_\perp$ one needs to select leptons
with individual transverse momenta greater than the maximum electron energy
in the rest system of the decaying open charm or bottom meson.
Typically such a threshold value is given by 1 (2) GeV for charm (bottom).
The results of our calculations of the $Q_\perp$ spectra for
RHIC conditions are displayed
in fig.~3a. We select here dileptons with transverse mass within the
interval 2 GeV $< M_\perp <$ 3 GeV. The value of the minimum transverse
momentum of leptons is chosen as $p_\perp^{\rm min} =$ 1 GeV.
The detector acceptance destroys the ideal $M_\perp$ scaling,
i.e., the thermal yield gets some structure reflecting the
$p_\perp$ cut and the geometry.
One can see the threshold
like behavior of the dilepton spectrum from charm decays:
there is a rapid dropping of the yield at $Q_\perp >$ 1.3 GeV.
On the contrary the thermal dilepton spectrum sticks out up to the kinematical
boundary $Q_\perp = M_\perp$. Unfortunately such a value of
$p_\perp^{\rm min} =$ 1 GeV is not large enough to cause a similar
suppression effect for leptons from bottom decays.
Nevertheless as pointed out in ref.~\cite{Vogt1} and as seen
in fig.~3a the contribution from bottom decays is suppressed
(this is due to the narrow rapidity gap covered by PHENIX)
but it is clearly above the thermal yield for the initial quark fugacity
$\lambda_q =$ 0.1.
To estimate the chances to see
a thermal dilepton signal over the background from bottom decays we
perform variations of the initial quark fugacity.
Due to the factor $\lambda_q^2$ in the thermal rate eq.~(9), even a
comparatively small increase of the initial quark phase space saturation
can result in a noticeable enhancement of thermal dileptons compared
to the decay background. This can be seen in fig.~3a for the
initial quark fugacities $\lambda_q =$ 0.5 and 1.
For the latter value the thermal signal is above the
bottom decay contribution.

For LHC conditions one can obtain the threshold like effect in the
$Q_\perp$ spectrum even for the dielectrons from the massive bottom
decays. In fig.~3b we show our results for dileptons within the
transverse mass interval 5.25 $< M_\perp <$ 5.75 GeV. The electron
transverse momentum cut $p_\perp^{\rm min} =$ 2 GeV causes a
sufficiently strong threshold effect both for charm and bottom, while the
dielectrons from thermalized matter exhibit the approximate $M_\perp$
scaling even for such a value of $p_\perp^{\rm min}$.
It gives therefore the opportunity to measure the thermal dilepton signal
on the basis of the $M_\perp$ scaling at LHC conditions.
Actually such measurements could be performed for lepton pairs with high
enough transverse momentum but smaller invariant mass.
In particular, as follows from fig.~3b, for a fixed value of $M_\perp \simeq$
5.5 GeV the proper region for transverse pair momenta is about
4 GeV $< Q_\perp <$ 5.5 GeV. This window increases considerably
with increasing initial quark fugacity $\lambda_q^i$ of thermalized parton
matter.

We perform also calculations of the $Q_\perp$ spectra at fixed $M_\perp$
for decay dileptons including the Peterson fragmentation
function \cite{Peterson}, where the function
\begin{equation}
f(z) = \left[
z \left(1 - \frac 1z -\frac{\epsilon}{1 - z} \right)^2
\right]^{-1}
\end{equation}
($\epsilon =$ 0.06 (0.02) for charm (bottom) \cite{Pet_par}) prescribes
the probability to change the spatial momentum of a heavy quark
from $\vec p$ to $z \vec p$ during
hadronization. Independently we also study the influence of energy
loss of heavy quarks in the deconfined medium.
Both these effects cause a suppression of the dilepton yield from
open charm and bottom, in particular in the small-$Q_\perp$
(large-$M$) region. Since the suppression effect due to Peterson's
fragmentation is quite small (typically within factor 2)
we do not display it separately.
On the contrary, the degradation of the heavy quark momenta due to energy
loss in the deconfined medium causes a considerably suppression
of the contribution of decay leptons in the $Q_\perp$ spectrum
at fixed value of $M_\perp$. In fig.~4 we show the results of our calculations
for LHC conditions with the same kinematical cuts as in fig.~3b.
As seen in this plot, both charm and bottom are considerably suppressed
due to energy loss effects, in particular in the region of smaller values
of $Q_\perp$.
These calculations exploit the model I.
In case of model II the suppression effect is much stronger.
With respect to the wanted thermal dilepton signal the suppression
of the $Q_\perp$ spectra from bottom decays appears most important.
The proper $Q_\perp$ window to extract thermal
dileptons becomes greater, e.g., 3 GeV $< Q_\perp <$ 5.5 GeV
for initial quark fugacity $\lambda_q =$ 0.5.
Therefore, the energy loss effect obviously helps to measure thermal
dielectrons with ALICE at LHC on the basis of the $M_\perp$
scaling behavior.

To clarify whether the proposed signal is experimentally feasible we
estimate the background from $\pi^0$ decays within the same
kinematical gates. For this background study we also perform Monte Carlo
simulations of $\pi^0 \to e^+ e^- \gamma$
decays utilizing the distribution
$dN_{\pi^0} / dy dp_\perp = (dN_{\pi^0}/dy) \exp \{p_\perp/T_\pi \}$
with $T_\pi =$ 200 MeV in accordance with present SPS results
for central Pb + Pb collisions at 158 AGeV beam energies
\cite{NA49}. The slope parameter $T_\pi$ is a convolution of local freeze-out
temperature and transverse flow. The latter one is expected to increase from
SPS to RHIC to LHC due to higher rapidity densities, and it might cause a
slightly higher value of $T_\pi$. We choose here the fictitious values
$dN_{\pi^0}/dy =$ 2500 (500) for LHC (RHIC). Each $\pi^0$ is
three-body decayed in its rest frame with the branching
ratio of 1.2\% \cite{PDB} and then the resulting $e^+ e^-$ pair is boosted
and filtered.
We find the individual $\pi^0$ decay dileptons with
5.25 GeV $< M_\perp <$ 5.75 GeV three orders of magnitude below the
thermal dilepton signal and only at $Q_\perp >$ 5.2 GeV.
The flattening of the pion spectrum at large values of $p_\perp$,
due the Cronin effect and hard scatterings, is not covered by our
parametrization of the $p_\perp^{\pi^0}$ spectrum;
it will enhance the pion decay contribution. Otherwise some quenching
\cite{Wang} at large values of $p_\perp^{\pi^0}$ can counteract and
balance such effects. The
finite experimental momentum resolution will cause a decay contribution also
at somewhat smaller values of $Q_\perp$.
However, these effects will hardly close the window
for the thermal signal.
The next-to-leading order Drell-Yan contribution is also estimated
to be strongly suppressed at larger values of $Q_\perp$
(say, for $Q_\perp >$ 1 GeV) because of the chosen large $p_\perp$ cut.
The most prominent danger however comes from
the combinatorical background from uncorrelated hadron decays:
Only after experimentally removing this
background together with the background from the conversion
dileptons, emerging from energetic photons in the detector material,
one has a chance to see the interesting signal.

Fully analog considerations also apply at RHIC. However, selecting
electron pairs with such large values of $M_\perp \approx$ 5.5 GeV
and choosing $p_\perp^{\rm min} =$ 2 GeV results in
prohibitory small counting rates which seem to make such a strategy
not feasible.
For the above choice of $M_\perp \approx$ 2.5 GeV
and choosing $p_\perp^{\rm min} =$ 1 GeV
the dileptons from $\pi^0$ decays produce here
a contribution which is three orders of magnitude above the thermal
signal, however only in the interval $Q_\perp >$ 1.95 GeV.
Therefore, one remains in a situation as depicted in fig.~3a and
discussed above.

\section{Conclusions}

In summary we analyze the chances of observing a thermal dilepton signal
in high-energy heavy-ion collisions.
Exploiting currently estimated energy loss effects of heavy quarks passing
through deconfined matter we find that the invariant mass spectra
of dileptons, as measurable in the detector facilities
PHENIX and ALICE, offer hardly a chance to find thermal dileptons
with $M >$ 2 GeV.
Even if in optimistic energy loss scenarios the dileptons from
correlated charm and bottom decays compete with the Drell-Yan background,
the thermal signal is significantly smaller. Since the energy loss
is a poorly quantified effect we stress the need to make attempts to
verify it, for instance via the single electron $p_\perp$ spectra which should
reflect the transverse momentum degradation of parent charm and bottom
mesons.

The double differential rate $dN_{e^+ e^-}/dM_\perp^2 \, dQ_\perp^2$
with $M_\perp$ in a narrow interval and with a suitable
$p_\perp^{\rm min}$ cut on the individual leptons,
however, seems to allow for a window
at large values of the pair transverse momentum
where the thermal yield shines out.
At RHIC this is probably not feasible,
while at LHC the estimated high initial temperature of the
parton matter causes a stronger thermal signal. The basic
issue is here whether the combinatorical background of many
other strong sources can be subtracted with sufficient accuracy.

\subsection*{Acknowledgments}

Stimulating discussions with
R. Baier, E.V. Shuryak, M. Tannenbaum, R. Vogt, X.N. Wang, and G.M. Zinovjev
are gratefully acknowledged.
O.P.P. thanks for the warm hospitality of the nuclear theory group
in the Research Center Rossendorf.
This work is supported in parts by BMBF grant 06DR829/1 and
USFFR grant 2.5.1/41.

\newpage


\begin{figure}
\centerline{{\psfig{file=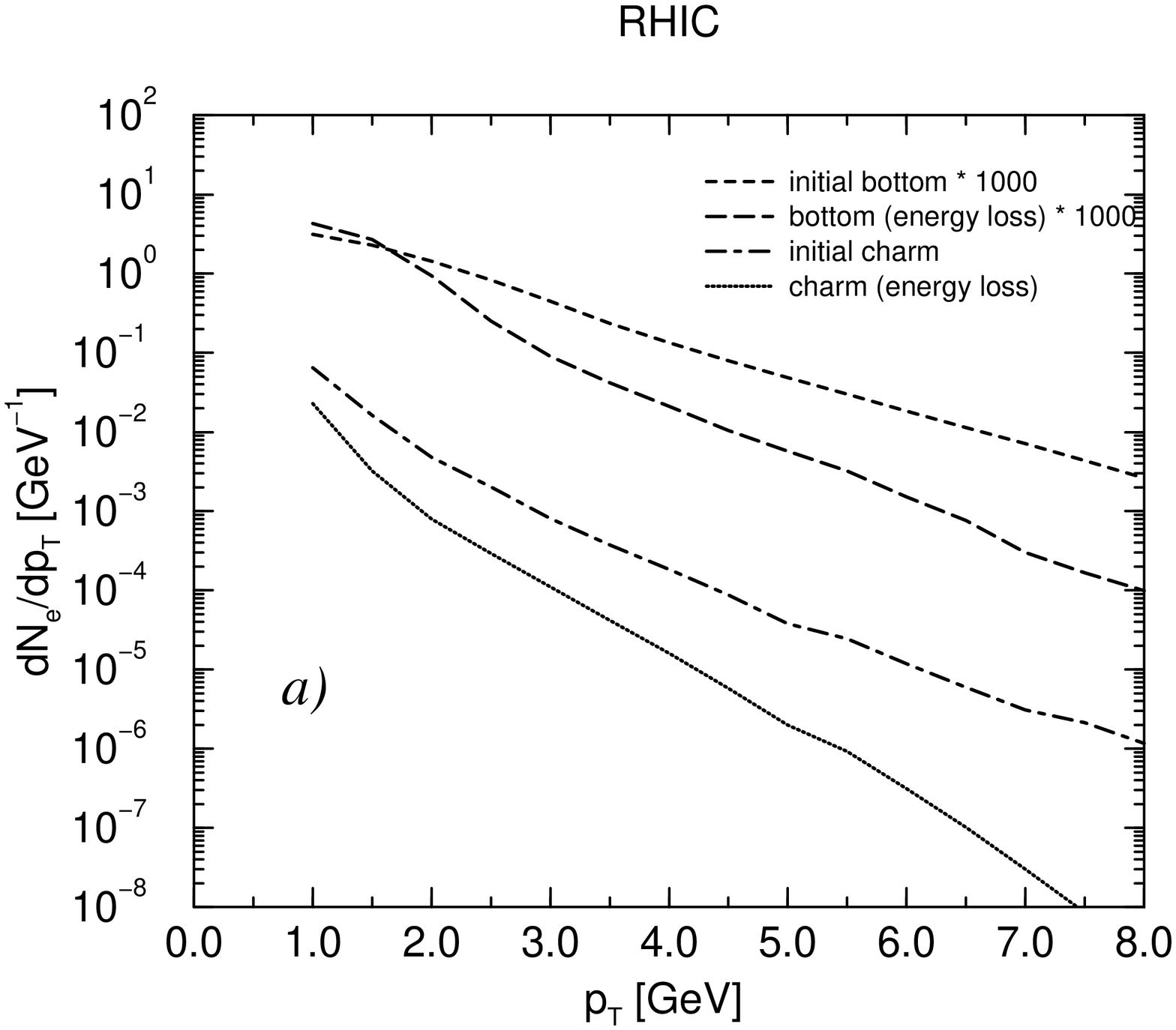,width=10cm,angle=-90}}}

\vspace*{6mm}

\centerline{{\psfig{file=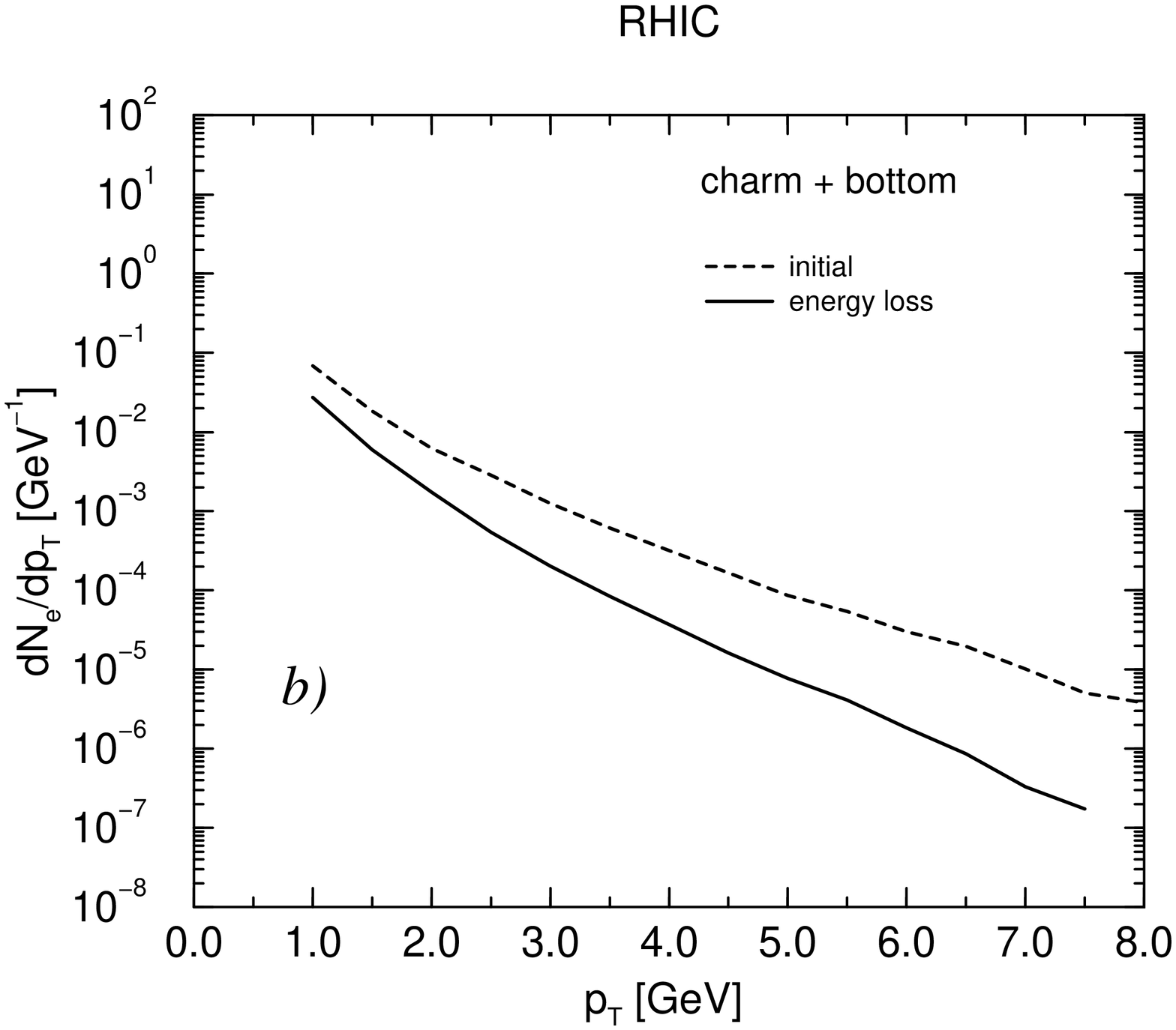,width=10cm,angle=-90}}}

\vspace*{6mm}

\caption{
The transverse momentum spectra of single electrons
from $D$ and $B$ meson decays (a) and their sum (b)
at RHIC energies within the PHENIX acceptance.
Displayed are the spectra before (''initial'') and after energy
loss within model II.
}\end{figure}
\newpage
\begin{figure}
\centerline{{\psfig{file=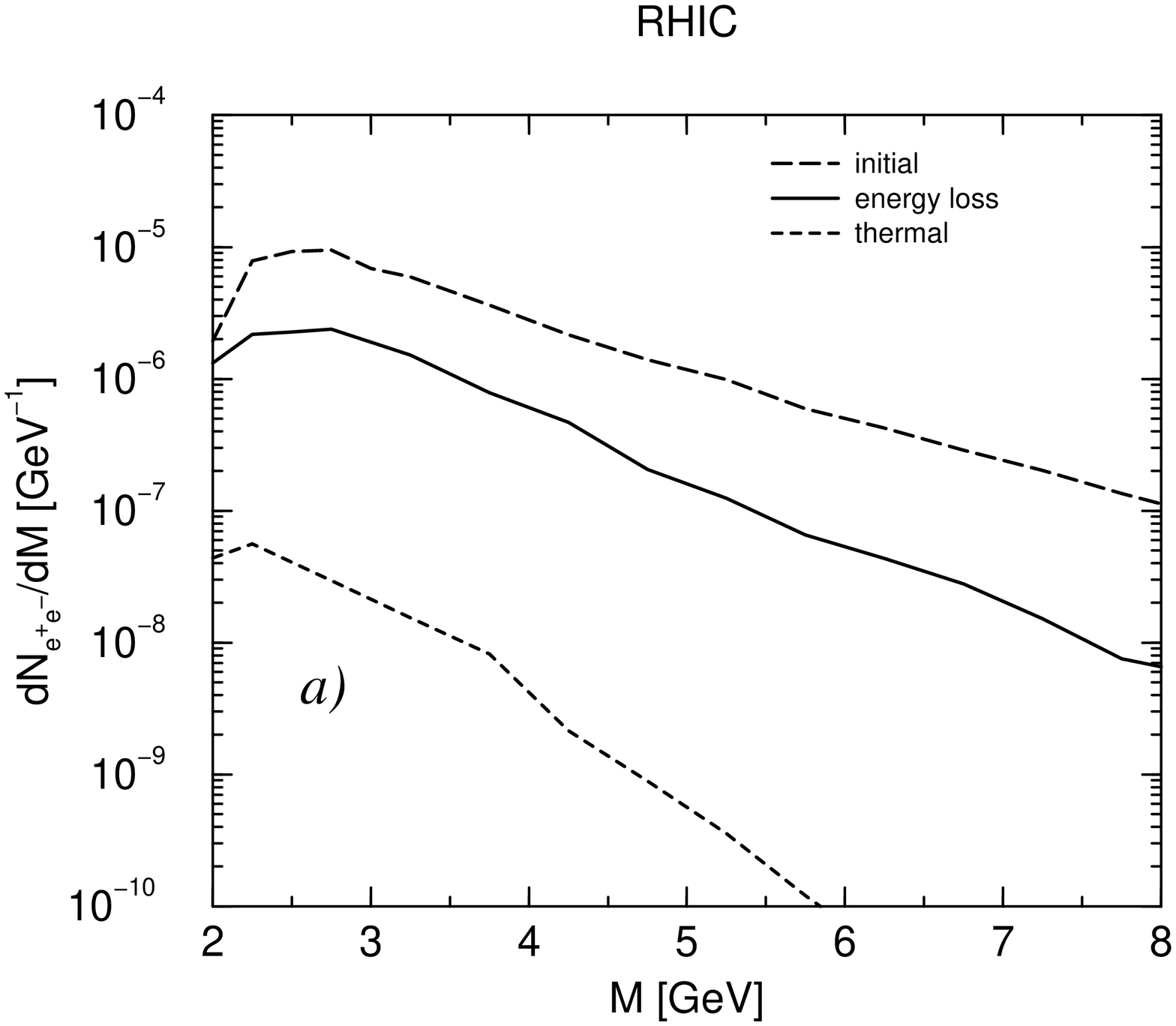,width=10cm,angle=-90}}}

\vspace*{6mm}

\centerline{{\psfig{file=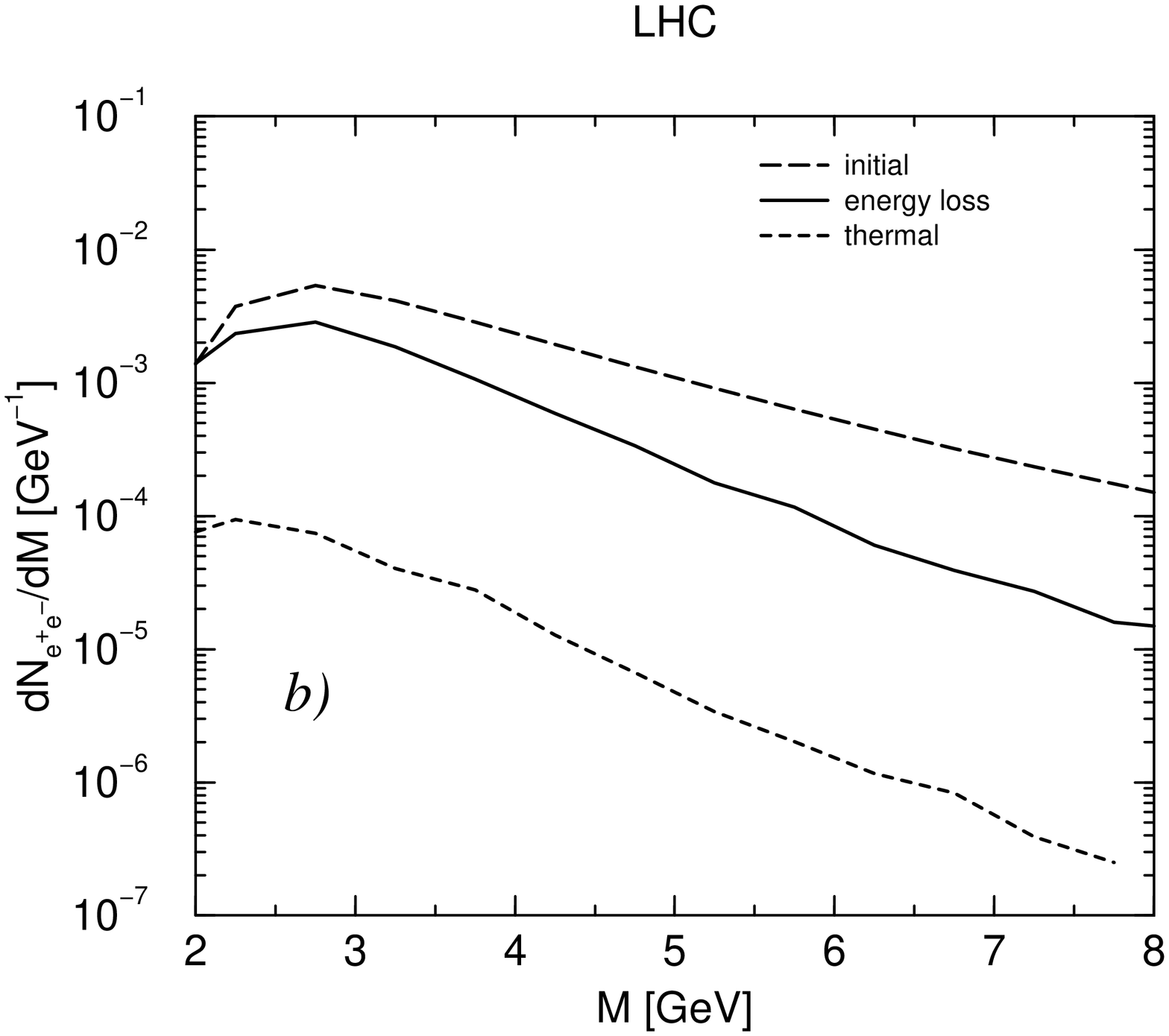,width=10cm,angle=-90}}}

\vspace*{6mm}

\caption{
Dielectron spectra from correlated charm and bottom decays
as a function of the invariant mass;
long-dashed (solid) lines are for results  without (with) energy loss
according to model II for RHIC with PHENIX acceptance (a)
and model I for LHC with ALICE acceptance but without high-$p_\perp$ cut (b).
The dashed lines depict the thermal dielectron spectrum.
For the strength of the Drell-Yan contribution cf.\
[9,10]
}\end{figure}
\newpage
\begin{figure}
\centerline{{\psfig{file=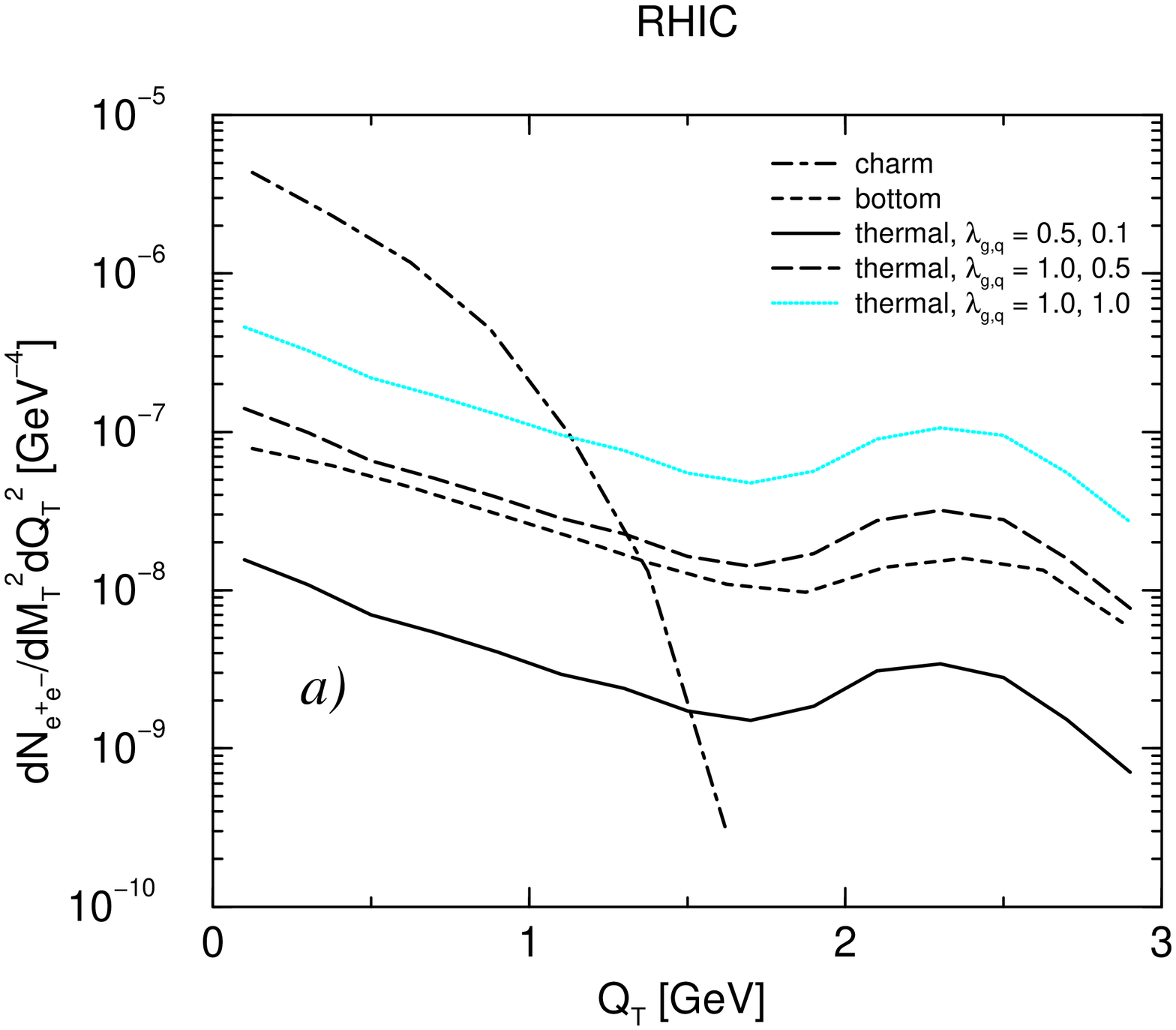,width=10cm,angle=-90}}}

\vspace*{6mm}

\centerline{{\psfig{file=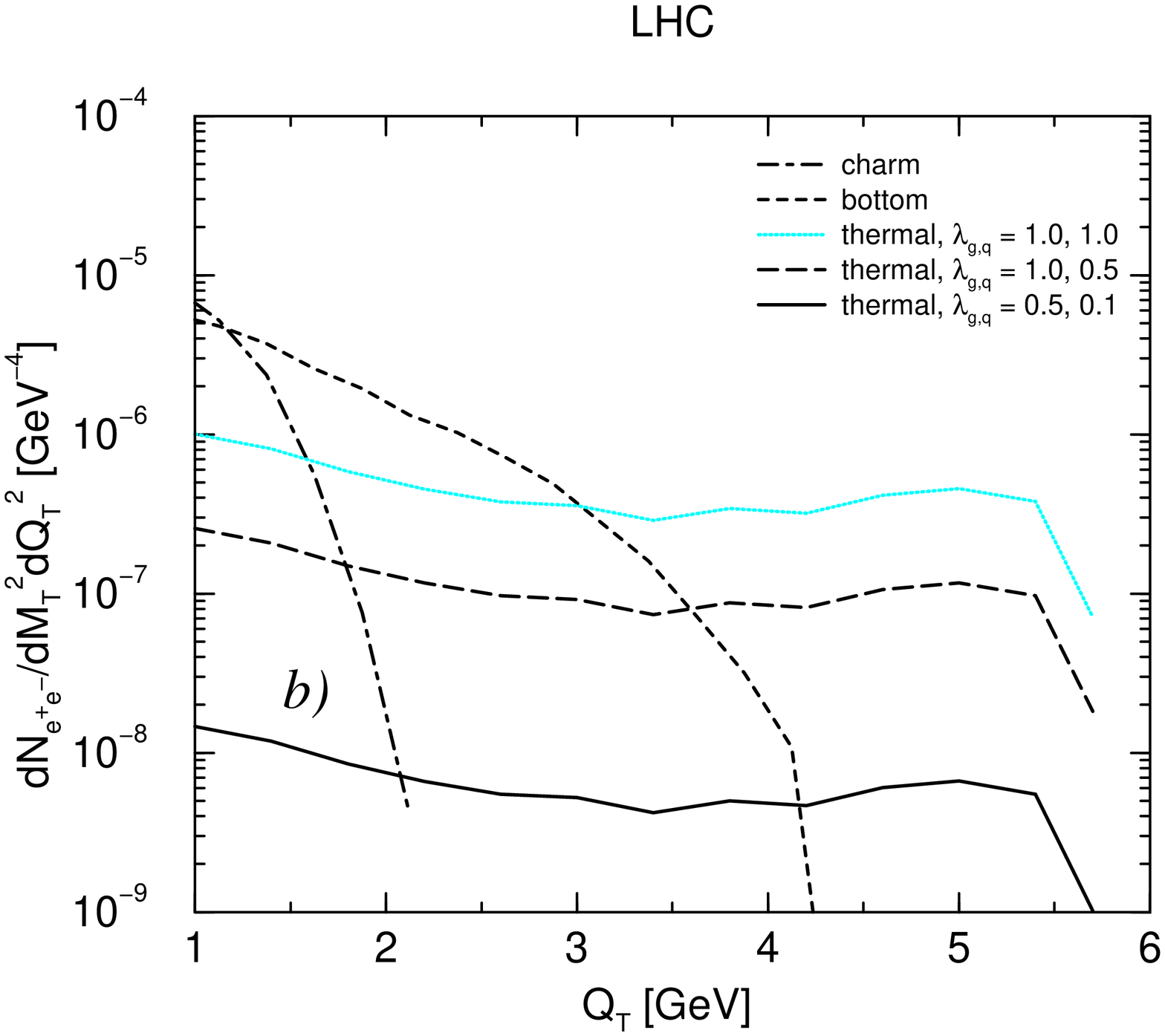,width=10cm,angle=-90}}}

\vspace*{6mm}

\caption{
The transverse momentum spectra of dielectrons at constrained transverse
masses:
(a) RHIC with PHENIX acceptance, $p_\perp >$ 1 GeV
and 2 GeV $< M_\perp <$ 3 GeV,
(b) LHC with ALICE acceptance, $p_\perp >$ 2 GeV
and 5.25 GeV $< M_\perp <$ 5.75 GeV.
The dash-dotted (short-dashed) lines are for correlated charm (bottom) decays.
The full, long-dashed and gray-dotted lines depict the thermal
yield for various initial quark and gluon fugacities.
}\end{figure}
\newpage
\begin{figure}
\centerline{{\psfig{file=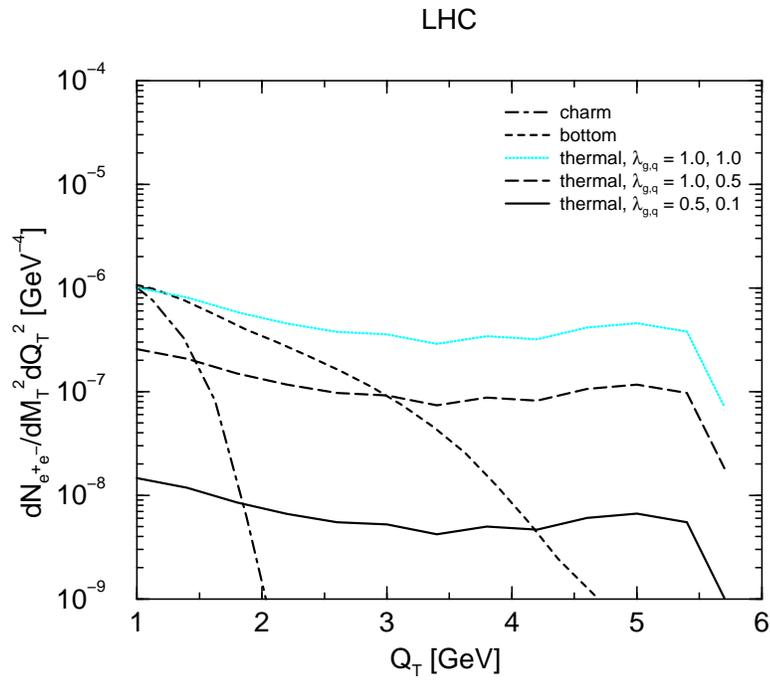,width=10cm,angle=-90}}}

\vspace*{6mm}

\caption{
The same as in fig.~3b but with energy loss effects according to
model I.
}\end{figure}

\end{document}